\documentclass[aps,pre,twocolumn,showpacs,superscriptaddress]{revtex4-1}
\usepackage{times}
\usepackage{amsmath}
\usepackage{natbib}
\usepackage{graphics}
\usepackage[dvips]{graphicx}
\usepackage{color}
\usepackage[breaklinks=true]{hyperref}
\usepackage{breakcites}
\usepackage{hypernat}
\usepackage{float}

\begin{document}

\title{Stochastic win-stay-lose-shift strategy with dynamic aspirations in evolutionary social dilemmas}

\author{Marco A. Amaral}
\email{marcoantonio.amaral@gmail.com}
\affiliation{Departamento de F\'\i sica, Universidade Federal de Minas Gerais, Caixa Postal 702, CEP 30161-970, Belo Horizonte - MG, Brazil}

\author{Lucas Wardil}
\affiliation{Departamento de Fisica, Universidade Federal de Ouro Preto, Ouro Preto, CEP 35400-000 MG, Brazil}

\author{Matja{\v z} Perc}
\affiliation{Faculty of Natural Sciences and Mathematics, University of Maribor, Koro{\v s}ka cesta 160, SI-2000 Maribor, Slovenia}
\affiliation{CAMTP -- Center for Applied Mathematics and Theoretical Physics, University of Maribor, Krekova 2, SI-2000 Maribor, Slovenia}

\author{Jafferson K. L. da Silva}
\affiliation{Departamento de F\'\i sica, Universidade Federal de Minas Gerais, Caixa Postal 702, CEP 30161-970, Belo Horizonte - MG, Brazil}

\begin{abstract}
In times of plenty expectations rise, just as in times of crisis they fall. This can be mathematically described as a Win-Stay-Lose-Shift strategy with dynamic aspiration levels, where individuals aspire to be as wealthy as their average neighbor. Here we investigate this model in the realm of evolutionary social dilemmas on the square lattice and scale-free networks. By using the master equation and Monte Carlo simulations, we find that cooperators coexist with defectors in the whole phase diagram, even at high temptations to defect. We study the microscopic mechanism that is responsible for the striking persistence of cooperative behavior and find that cooperation spreads through  second-order neighbors, rather than by means of network reciprocity that dominates in imitation-based models. For the square lattice the master equation can be solved analytically in the large temperature limit of the Fermi function, while for other cases the resulting differential equations must be solved numerically. Either way, we find good qualitative agreement with the Monte Carlo simulation results. Our analysis also reveals that the evolutionary outcomes are to a large degree independent of the network topology, including the number of neighbors that are considered for payoff determination on lattices, which further corroborates the local character of the microscopic dynamics. Unlike large-scale spatial patterns that typically emerge due to network reciprocity, here local checkerboard-like patterns remain virtually unaffected by differences in the macroscopic properties of the interaction network.
\end{abstract}

\pacs{89.75.Fb, 87.23.Ge, 89.65.-s}
\maketitle

\section{Introduction}
\label{Introduction}

Cooperation has been theoretically studied in evolutionary game theory \cite{MaynardSmith1982a, Weibull1995, Hofbauer1998, Mesterton-Gibbon2001, Nowak2006}. The prisoner's dilemma game \cite{Axelrod1984, Nowak2006} is classically used to portrait the situation where it is beneficial for all members of a group to cooperate but, whatever the others do, it is always more beneficial for the individual do defect. Due to its broad interpretation, prisoner's dilemma inspired research in social and natural sciences alike \cite{fudenberg_e86, nowak_n93, santos_prl05, Wu2005, Imhof2005, Fu2007, wu_zx_pa07, Gomez-Gardenes2007, Tanimoto2007, poncela_njp07, Poncela2009, Antonioni2011, Tanimoto2012a, Gomez-Gardenes2012, szolnoki_pre14, Tanimoto2013, tanimoto_amc15}. In the prisoner's dilemma game, two players can either cooperate (C) or defect (D). Mutual cooperation yields a payoff $R$ (reward) and mutual defection yields $P$ (punishment). If players have different strategies, the defector receives $T$ (temptation) and the cooperator receives a small payoff $S$ (sucker). Usually prisoner's dilemma follows the hierarchy $T>R>P>S$ \cite{Nowak2006, Szabo2007, perc_bs10}. In classical game theory, defection is the Nash equilibrium and, therefore, the rational choice. Even so cooperation flourishes in human societies, between members of the same species and in some inter-species symbiosis \cite{wilson_71, Skutch1961, Nowak2011a}.

In an original approach to evolutionary game theory, Nowak studied spatially distributed populations where players copy the fittest strategy. This model showed how cooperation can exist in a sea of defectors, the so called ``spatial reciprocity'' mechanism -- cooperators spontaneously form clusters where they support each other. After this,  new mechanisms of cooperation promotion were investigated, usually based on some kind of reciprocity. Among the most studied mechanisms are kin selection \cite{Hamilton1964}, direct and indirect reciprocity \cite{trivers_qrb71, axelrod_s81}, network reciprocity \cite{Nowak1992a, wardil_epl09, wardil_pre10},  group selection \cite{wilson_ds_an77} and heterogeneity \cite{perc_bs10, Amaral2016, Amaral2015}. The most common interpretation of these models are in terms of biological evolution, using  birth-death-like dynamics to model the replication of strategies \cite{schuster_jtb83, Sysi-Aho2005,  nowak_s04, MaynardSmith1982a}: Players with higher payoff grow in population, alike the selection of the fittest \cite{MaynardSmith1982a,Nowak2006, Szabo2007}.
Dynamics where players only copy strategies previously available in the system are called \textit{non-innovative dynamics}. On the other side, in \textit{innovative dynamics}  new strategies can arise, for example, via mutation, exploration, testing etc \cite{Nowak2006, Szabo2007, wang_s_ploso08, roca_epjb09}. While non-innovative dynamics usually describe long term evolutions, usually innovative dynamics represent situations where players can take cognitive responses to the environment, like human interactions \cite{vukov_njp12, blume_a_geb10, Szabo2007, bonawitz_cg14}.

Recent works support the idea that human interactions are strongly influenced by cognitive choices other than just copy mechanisms \cite{Dalton2010, vukov_njp12, macy_pnas02, blume_a_geb10, bonawitz_cg14}. New behavior can emerge and people often change their opinions without the need of a ``copy source''. In contrast to simpler species, where the evolution of strategies is basically governed by birth-death processes, in the human species strategies spread via other mechanisms as well \cite{Sysi-Aho2005, Szabo2007, grujic_pone10,szolnoki_epl11}. For example, recent experimental results show that individuals decided the strategy in the next round in moody way\cite{Wedekind1996}, that is, individuals will cooperate in the next round if they have cooperated in the previous round, otherwise defection will follow.  Also, experimental evidence shows that individual decisions are guided by aspiration levels \cite{Dalton2010}. So it is typical of human behavior to adopt innovative dynamics. Note that, in game theory, all possible strategies are already defined in the strategy space; hence, innovation refers to the possibility of a new strategy emerge in a monomorphic population. With this as motivation, we explore innovative dynamics in the context of evolutionary game theory.

The Win-Stay-Lose-Shift strategy (WSLS; also know as Pavlov) is an innovative strategy that relies on cognitive capabilities, instead of replicating process \cite{Thorndike1911, kraines_td89, nowak_n93, vukov_njp12, bonawitz_cg14}. WSLS was proposed in the famous Axelrod Tournaments \cite{Axelrod1984, Nowak2006} and proved to be very efficient against others strategies in the iterated prisoner's dilemma in infinite, well-mixed populations. It performed similarly to the famous Tit-For-Tat and fare even better in noise environments \cite{nowak_n93}. A WSLS player keeps its strategy if its payoff is above a desired level --the individual aspiration-- and changes when it is bellow the aspiration. After the initial success in Axelrod tournament, many different WSLS-like strategies  have been proposed \cite{yang_hx_pa12, chen_xj_pre08, chen_xj_pa08, palomino_ijg99, macy_pnas02, Oechssler2002, chen_jsm15}. Nevertheless, the aspiration level is usually implemented as a global external parameter and the decision-making process  is deterministic: players always change strategy if  payoff is bellow the aspiration level.

Inspired by recent works that interpret aspiration as dynamic, or co-evolving parameter \cite{perc_bs10, perc_pone11, pacheco_prl06, pacheco_jtb06, Liu2012, posch_prslb99, platkowski_aml09, chen_jsm15, Wang2010, cho_jet05}, we propose a model where an individual aspiration  is dynamically determined by the average payoff of its neighbors. In this way there is a variety of different aspirations that evolve spatial and temporally in the population. Specifically, this model relates to the rationale that aspiration levels tend to follow the wealth level of one's society. In the middle of a crisis, humans tend to lower what they expect to receive in interactions. On the other side, it is normal to want a higher payoff when all your peers are faring better than you. We note that previous works on WSLS spatial games usually considered populations where WSLS is one strategy, among others, that can be transmitted via copying mechanisms \cite{Posch1999, roca_epjb09, Szabo2007}. Here we consider a population where only two strategies are available -- cooperation and defection -- and the update rule is defined as a win-stay-lose-shift behaviors: if my payoff  is bellow the average payoff of my neighbors then I change my strategy, otherwise I keep my current strategy.  Since new strategies can emerge in monomorphic populations, this dynamics is innovative.

Our main objective here is to analyze the proposed model, comparing it to the classic version of non-innovative dynamics, highlighting what differences can arise. In the next section we define our model precisely. We also define the imitation rule, a non-innovative dynamics that is well studied in the literature, that will serve as a baseline for comparison. In Results, we study the master equation and its implications for the WSLS model, as well as the Monte-Carlo numerical simulations in square and scale-free networks. We considered the most studied two-players dilemma games -- prisoner's dilemma, snow-drift and stag-hunt games.  Finally,  we summarize our results in  the Conclusion section.

\section{Mathematical model}
\label{Model}

Players have only two possible strategies: cooperation (C) and defection (D). Individuals are represented by the nodes of a network and the game happens on pairwise interactions between players and their neighbors. In each interaction, players receive a payoff according to the usual payoff matrix \cite{ Szabo2007, Nowak2006}:

\[
 \bordermatrix{~ & C & D \cr
                  C & 1 & S \cr
                  D & T & 0 \cr},\]
where $T\in[0,2]$ and $S\in[-1,1]$.  Note that the parametrization $G=(T,S)$ spans four different classes of games: prisoner's dilemma (PD),  snow-drift  (SD),  stag-hunt  (SH), and  harmony games (HG)  \cite{Szabo2007, Nowak2006,perc_bs10}.

The evolution of strategies is defined in  two phases. First, players interact with their neighbors and accumulate the payoff obtained in each interaction. Second, players may change their strategy according to an update rule. We study in this work the win-stay-lose-shift with dynamic aspiration update rule and compare it to the classic imitation update rule.

\subsection{WSLS with dynamic aspiration}

In this update rule individuals change their strategies depending on the degree of satisfaction with their current payoff in comparison to the average payoff of their neighbors. The WSLS strategy  is usually defined in terms of fixed aspiration level as an external parameter \cite{yang_hx_pa12, chen_xj_pre08, chen_xj_pa08, palomino_ijg99, macy_pnas02, Oechssler2002}. Recently, papers started using heterogeneous and time evolving aspiration levels \cite{Wang2010, perc_bs10, perc_pone11, pacheco_prl06, pacheco_jtb06, Liu2012, posch_prslb99, platkowski_aml09, chen_jsm15, cho_jet05}. Our model merges a probabilistic decision-making process with the concept of aspiration as the average payoff of the neighbors. In accordance with other coevolutionary models \cite{perc_bs10, perc_pone11, pacheco_prl06, pacheco_jtb06, platkowski_aml09, Wang2010}, we intend to make the aspiration an emerging property, intrinsic to the system. At every time step, a player is randomly selected to update its strategy. A player, $i$,  changes its strategy --  a cooperators changes to defection and a defector changes to cooperation -- with  probability
\begin{equation}\label{wslseq}
p(\Delta u_{i})=\frac{1}{ 1+e^{-( \bar{u}-u_{i})/k} },
\end{equation}
where $\bar{u}$ is the average payoff of player $i$'s neighbors. This probability distribution, which is based on the Fermi-Dirac distribution of statistical physics, is widely used in evolutionary dynamics \cite{Hauert2005}. The parameter $k$ measures how often players  make ``irrational'' choices, changing strategies against the rationality prescribed by the model \cite{Szabo2007}. In the literature,  usually we find $k \in[0.001,0.3]$ to simulate a small, but non-zero, chance of a player making mistakes (trembling hand) \cite{Szabo2007, Nowak2006}.

The WSLS update rule has several distinct features.
First, the aspiration value  is not an external parameter; it is an emerging property of the system.
Second, each site  has its own aspiration value.
Third, the aspiration is subject to temporal and spatial variations.
It seems natural to determine the aspiration in terms of the neighborhood average, as people tend to lower their expectations during some global crises, while they raise the expectations when neighbors are faring better.

It is important to stress that we still have only two different strategies in the population C or D. Differently from usual works\cite{szolnoki_srep14}, here WSLS is not considered a ``pure'' strategy, rather it is an update mechanism.

\subsection{Imitation update rule}

As a baseline for comparison, we are going to contrast our model to the imitation update rule. In this rule, player $i$ update its strategy by randomly choosing one of its neighbors, $j$, and then comparing their payoffs. Site $i$ adopts the strategy of $j$ with probability
\begin{equation}\label{imitateeq}
p(\Delta u_{ij})=\frac{1}{ 1+e^{-(u_{j}-u_{i})/k} },
\end{equation}
where $u_{i}$ is the cumulative payoff of site $i$.

The imitation rule  is a non-innovative dynamic \cite{Szabo2007, Nowak2006}, because a player can only change its strategy to the  available ones in the population. This means that new strategies can never appear once extinguished (the system has absorbing states) and, most importantly, players never ``explore'' new strategies \cite{roca_epjb09, Szabo2007, Nowak2006}. This update rule is thus associated with the replication dynamics \cite{Szabo2007, Nowak2006,nowak_s04, MaynardSmith1982a} found in biological systems. The process of imitation is equivalent to local competition where death is a random, uniform process and reproduction rates are determined by the payoffs (fitness). In this context, without mutation, extinct species never re-appear.

\section{Results and discussion}

It is well known that there is a phase transition in the fraction of cooperation in square lattices with the imitation update rule: cooperation cannot survive for certain payoff parameters.  In the weak prisoner's dilemma \cite{Nowak2006,Szabo2007}, $P=S=0$, cooperation goes to extinction above a critical value of $T$. In contrast, we found that in the WSLS with dynamic aspiration cooperation always survives, even for large $T$. To understand this result, first we are going to study the master equation for the weak prisoner's dilemma in square lattices. We obtained analytical results for the limits of large $T$ and $k\rightarrow \infty$ and numerical results for general $T$ and $k$.  After this initial analysis, we are going to simulate the evolution of strategies in the entire parameter space, as well as in scale-free topologies.

\subsection{Master Equation}

On a square lattice, each player interacts with its four nearest neighbors.  A focal site, $f$,  can be in two states: cooperation or defection. In a mean field approximation we set the probability that the focal site is a cooperator equal to the population fraction of cooperators, $\rho$. Therefore
\begin{equation}\label{master}
\dot{\rho}= (1-\rho) \Gamma_{+(D\rightarrow C)} - \rho \Gamma _{-(C\rightarrow D)},
\end{equation}
where $\Gamma{+}$ ($\Gamma{-}$) is the transition rate accounting the probability that the focal player will change its strategy to C (D), given that its current strategy is D (C). We first use the simpler version of the master equation, the well-mixed approach. Here we consider that every player is connected to every other player. By doing so there is no spatial structure and we can consider the average payoff of a single cooperator, $u_c$ (defector, $u_d$), as the average payoff of all cooperators (defectors) in the population. Moreover, the average aspiration simply becomes the population average payoff ($\bar{u}$). The ODE to be numerically solved uses the transition rates:
\begin{eqnarray}
\Gamma_+= \frac{1}{1+e^{-(\bar{u}-u_d)/k}}, \\
\Gamma_-= \frac{1}{1+e^{-(\bar{u}-u_c)/k}}.
\end{eqnarray}

Assuming the weak prisoner's dilemma, $R=1$ and $S=P=0$, this gives us:
\begin{eqnarray}
u_d= && \rho ^2 R +(1-\rho)^2 P +\rho(1-\rho)(T+S)= \nonumber \\
= && \rho ^2 +\rho(1-\rho)T,  \\
u_c= && \rho R+(1-\rho)S = \rho R, \\
\bar{u}= && \rho T+(1-\rho)P = \rho T ,
\end{eqnarray}
and finally the ODE:
\begin{equation}
\frac{d \rho}{d t}=  \frac{1-\rho}{1+e^{-(\rho^2(1-T))/k}} -  \frac{\rho}{1+e^{-(\rho(\rho-1)(1-T))/k}}.
\end{equation}

\begin{figure}
\centering{\includegraphics[width = 8cm]{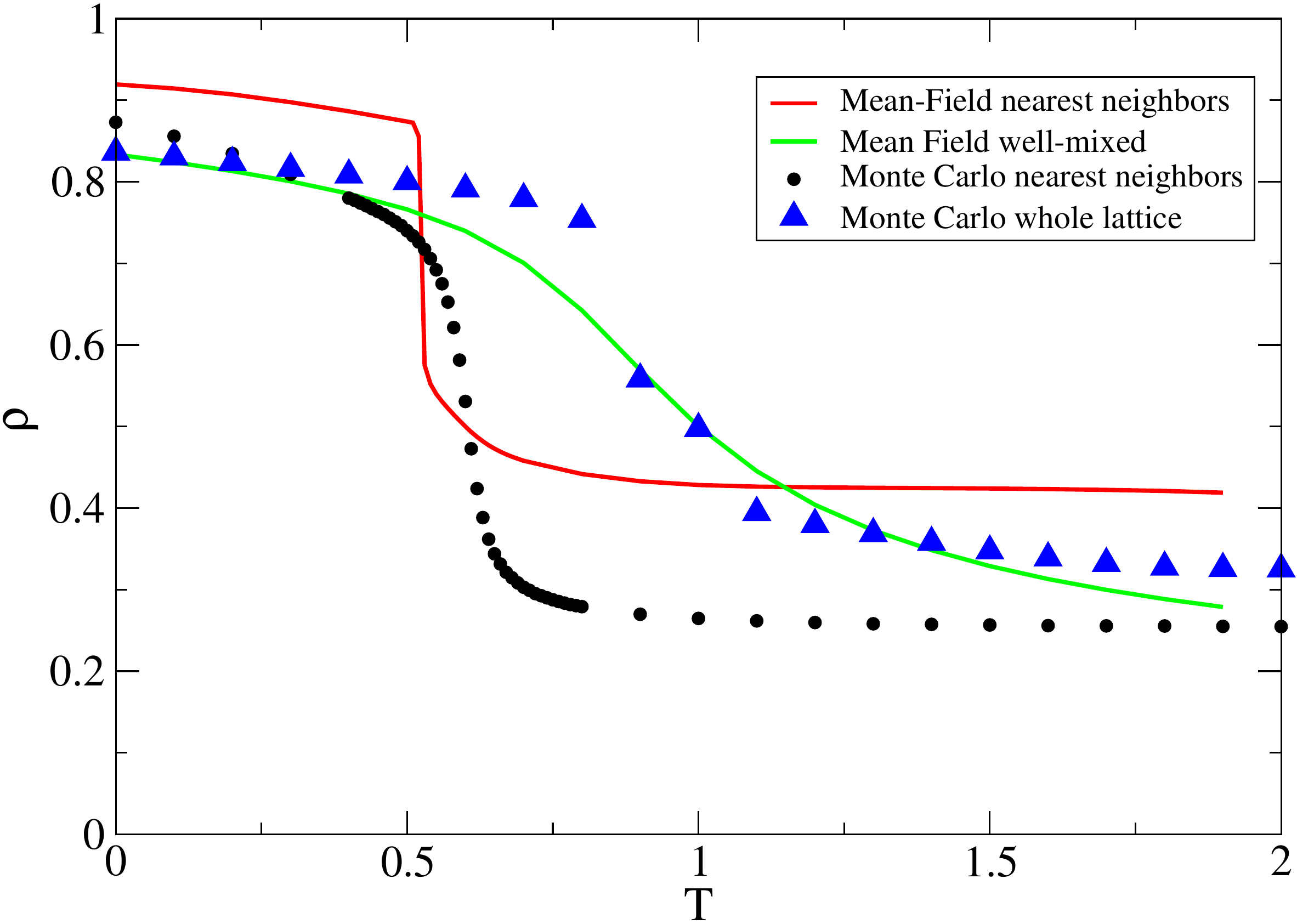}}
\caption{Equilibrium cooperation level in dependence on $T$ in the WSLS model, as obtained with Monte Carlo simulations and the integration of the master equation (see legend). ``Monte Carlo whole lattice'' refers to the aspiration being equal to the average payoff of the whole lattice rather than just the nearest neighbors, thus representing results for the two limiting cases concerning the interaction range of individual players. There are small quantitative differences between the presented curves. But more importantly, we see that the inflection point of the curves and the minimum level of cooperation are both very similar in all cases, thus pointing to a good qualitative agreement and a deeper mechanism promoting this effect.}
\label{fig_wslsnum}
\end{figure}

This first approximation relates to the case where there is no spatial structure, and therefore sites cannot rely on spatial correlation effects. The results for this model are shown in Fig.~\ref{fig_wslsnum}, together with the results for other studied cases. The next step if one wishes to account for spatial effects is to consider the nearest neighbor approximation \cite{Szabo2007, matsuda_h_ptp92, schuster_jtb83}. Here we use a focal site $i$ and define its chance of turning into a cooperator or defector. We do so by calculating exactly its payoff with the first four nearest neighbors of the square lattice. To calculate the transition rates, we consider all combinations of cooperators and defectors in the neighborhood of the focal site. The transition rates in the first neighbor approximation then become:
\begin{equation}\label{gamma}
\Gamma_{\pm}= \sum_{n=0}^{4} {{4}\choose{n}} \rho^n (1-\rho)^{(4-n)} P_{\pm}(u_f,\bar{u}),
\end{equation}
where $n$ is the number of cooperative neighbors for each configuration. The term ${{4}\choose{n}}$ is the binomial coefficient accounting different combinations of $n$ cooperators and $4-n$ defectors in a given configuration (for example, the configuration {CDDD} can repeat itself in 4 different ways, while {CCCC} happens only once). The term $\rho^n (1-\rho)^{4-n}$ weights the probability of such configuration, with $n$ cooperators, to happen. Finally $P_{\pm}(u_f,\bar{u})$ is the probability, in a given specific configuration, that the focal site will turn into a cooperator ($P_{+}$) or a defector ($P_{-}$). This probability is the only term that is directly dependent on the update rule chosen (Imitation or WSLS). Note that $u_f$ and $\bar{u}$ depend on the configuration. Since the solution for the master equation of the imitation model can be found in the literature \cite{Szabo2007, Nowak2006}, here we focus on the solution of the WSLS model.

The focal site compares its payoff, $u_f$, with the average payoff of the four neighbors. In a configuration where there are $n$ cooperators, the focal payoff is $u_f=nT$. To calculate $\bar{u}$, we have to estimate the payoff of each neighbor first. Let us  assume that the probability that a second neighbor of the focal is a cooperators is also equal to $\rho$. Given that the focal is a defector, a cooperative neighbor receives $0$ in the interaction with the focal and receives $3\rho$ on average in the interaction with the second-neighbors. In the same way, a defective neighbor receives $3 T \rho$ on average. Thus, the average payoff of the neighbors, $\bar{u}$, in any configuration with $n$ cooperators is
\begin{equation}
\bar{u_{\Omega}}=\frac{1}{4}\left[ n 3\rho+(4-n)3T\rho\right]= \frac{3}{4}(n\rho +4T \rho-nT\rho).
\end{equation}
Thus
\begin{equation}
P_{+}=\frac{1}{1+e^{-[T(12 \rho -3n\rho -4n)+3n\rho]/4k}}
\end{equation}
and
\begin{equation}
P_{-}=\frac{1}{1+e^{ [T(3n\rho +n-4-12\rho)-3n\rho+3n]/4k}}.
\end{equation}
The equilibria of the master equation can be found analytically in the limit  $k\rightarrow 0$, where players are assumed fully rational. In this limit, the fermi probabilities become
\begin{equation}
P_{\pm} = \begin{cases}
1 &\text{if $\bar{u}-u_{f}>0$}\\
1/2 &\text{if $\bar{u}-u_{f}=0$}\\
0 &\text{if $\bar{u}-u_{f}<0$}
\end{cases}
\end{equation}
and the transition rates will be just polynomial functions. In the limit of large $T$, the difference $\bar{u}-u_{f}$ can be explicitly evaluated for each neighborhood configuration of $n$ cooperators.  Solving this polynomials we found that
\begin{equation}
P_{-} = \begin{cases}
1 &\text{if $n<3$}\\
0 &\text{if $n=0$ and $\rho \neq 1$.}
\end{cases}
\end{equation}
and
\begin{equation}
\label{eq49}
P_{+} = \begin{cases}
1 &\text{if $n=0$}\\
0 &\text{if $n=1$ and $\rho<4/9$}\\
0 &\text{if $n>1 $,}
\end{cases}
\end{equation}
Simplifying the master equation (\ref{master}), in the limit for $k\rightarrow 0$ and large T, we get
\begin{equation}
\begin{split}
\dot{\rho}=(1-\rho)^5-\rho(1-\rho)[ (1-\rho)^3+4\rho(1-\rho)^2 \\
 +6\rho^2(1-\rho) +3\rho]
\end{split}
\end{equation}
This equation has a stable fixed point at $\rho^*\approx0.209$, which means that cooperation can coexist with defectors even for high temptation values. This is an interesting result in terms of cooperation survival and it goes along with different approaches on innovative dynamics \cite{roca_epjb09, szolnoki_pre14, Fort2005}. Note that, in equation \ref{eq49}, $P_+=0$ for $n=1$ only if $\rho<4/9$. Since $\rho^*<4/9$, the analysis is consistent.

We proceed with the numerical integration of the original master equation with arbitrary parameters (any $T$ value), using the 4th order Runge-Kutta method. Figure~\ref{fig_wslsnum} summarizes the results for both analytical equations (well-mixed population and nearest neighbors of the square lattice). For comparison, we also show the results of Monte Carlo simulations, which will be discussed more thoroughly in the next Section. We note that in Fig.~\ref{fig_wslsnum} ``Monte Carlo whole lattice'' refers to the aspiration of each site being equal to the average payoff of the whole lattice. It can be observed that the interaction topology slightly changes the results in both analytical and numerical models quantitatively, but not the main characteristics of the WSLS update rule. We also ran Monte Carlo simulations for different aspiration level ranges, varying it from just the four nearest neighbors to the whole lattice in a continuous fashion, and the results all fell between the two depicted limiting cases (the four nearest neighbors and the whole lattice) in Fig.~\ref{fig_wslsnum}. Looking at the results, we can observe that the WSLS update rule yields specific but generally valid results. Namely, there is always a minimum level of cooperation in the population even for large $T$ values, and a smoother decline in cooperation as $T$ increases when compared to the relatively steep and sudden transitions observed previously in imitation models (see also Fig.~\ref{fig_comp2D}). These results are thus intrinsically different from those obtained with imitative dynamics, even in the well-mixed case and regardless of the interaction range for the determination of payoffs and aspirations. We argue that this is due to the intrinsic micro-mechanism present in innovative dynamics, which if of course not present in the imitation model. We will further explore these mechanisms in the next Section with an analysis of the corresponding spatial patterns. Lastly concerning the results presented in Fig.~\ref{fig_wslsnum}, we also point out that the Monte Carlo simulations and the numerical solutions of the master equation agree very well qualitatively.

\begin{figure}
\centering{\includegraphics[width = 8cm]{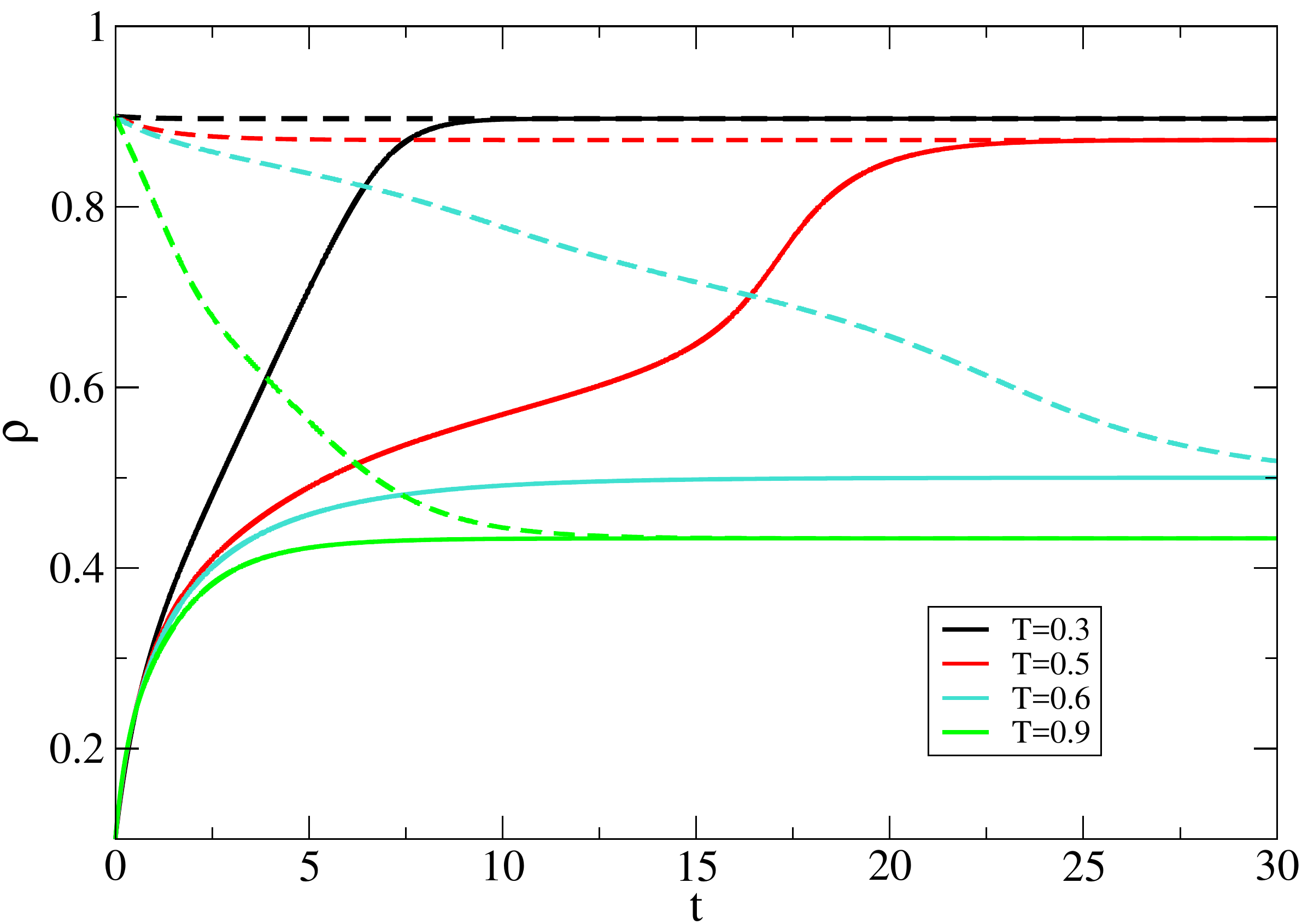}}
\caption{Time evolution of the average cooperation level in the mean-field approximation. The graph shows two sets of curves, each obtained with different initial conditions. The system always reaches a stable state, independent of the initial condition for each value of $T$.}
\label{fig_fwslsnum}
\end{figure}

Figure~\ref{fig_fwslsnum} shows that the system reaches a stable state independently of the initial fraction of cooperation. This is a very important feature of the proposed model, since it is well know that not every update rule will have an equilibrium state that is independent of the initial conditions \cite{roca_epjb09,Wang2012, Vainstein2001, arapaki_pa09}.

The mean-field technique is a good approximation to obtain insights and confirm  predictions. However, it often does not return the same results of the simulation in the structured population, some times not even qualitatively  \cite{Szabo2007, szabo_pre05}. In our case, it is interesting to notice that the mean-field approximation correctly predicts the existence of the minimum cooperation level.

\subsection{Monte Carlo simulations}

We use the asynchronous Monte Carlo procedure to simulate the dynamics. A random player, $i$, is selected and the cumulative playoff of $i$, as well as the payoff of the first and second neighbors of $i$, are calculated. Then player $i$ decides to change its strategy based on the update probability \ref{wslseq} for WSLS or \ref{imitateeq} for imitation dynamics.  One Monte Carlo step (MCS) consists of this process being repeated until each player has the change to change its strategy. We used $k=0.1$ in all simulations. For a detailed discussion on Monte Carlo methods in evolutionary dynamics see \cite{landau_00, binder_88, binder_rpp97, Szabo2007}.
In our  simulations we ran the algorithm until the system reaches an equilibrium state ($10^4-10^5$ iterations) \cite{Szabo2007, Huberman1993a}. Then we take the averages over 1000 Monte Carlo Steps (MCS) for $10-20$ different initial conditions. We use $10^4$ individuals distributed in a square lattice, unless stated otherwise. The square lattice have periodic boundary conditions and we start with homogeneous strategy distribution (we note that for our model the initial distribution did not change the final outcome).

\begin{figure}
\centering{\includegraphics[width = 8cm]{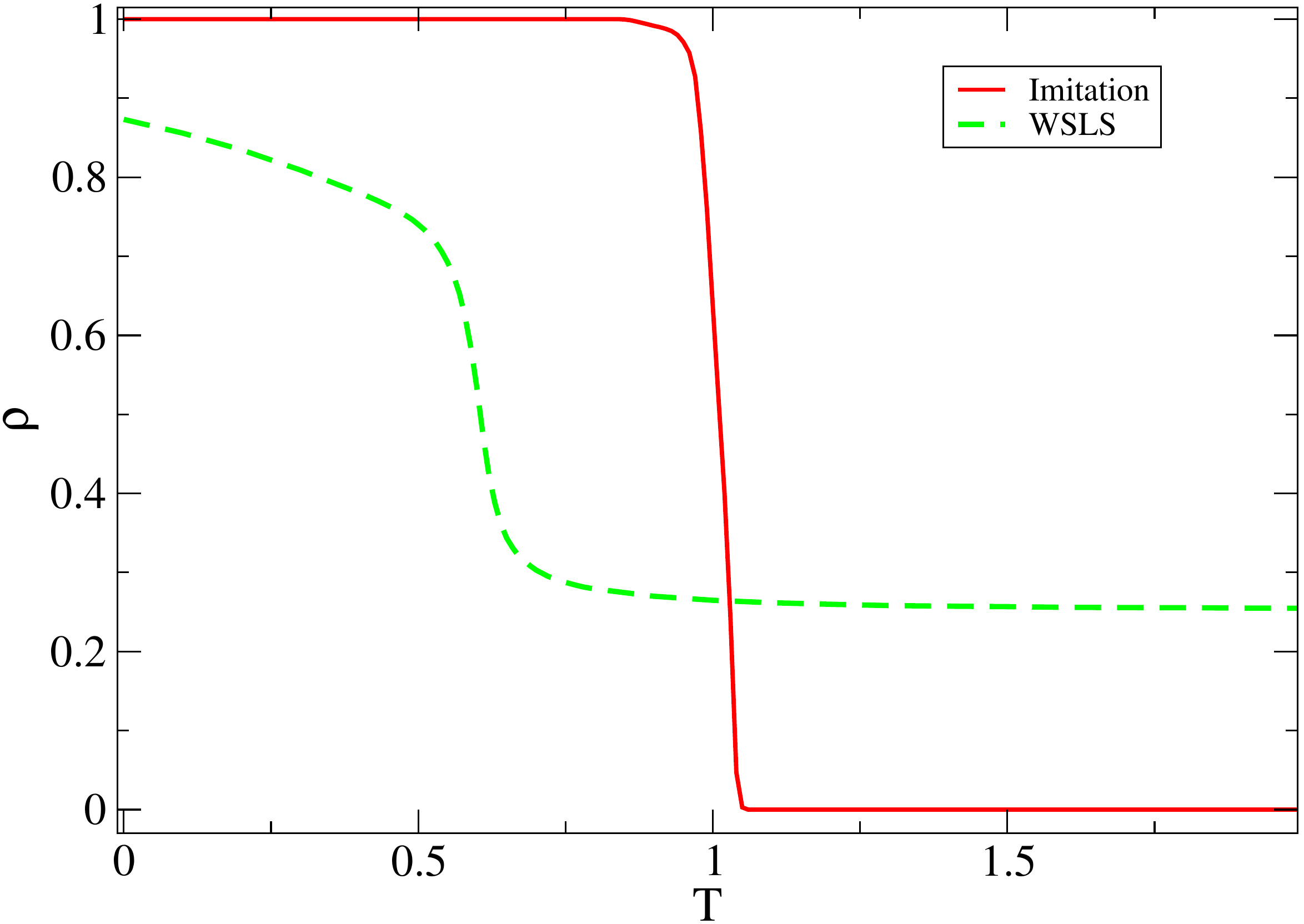}}
\caption{Fraction of cooperator as a function of $T$, as obtained with Monte Carlo simulations. We used the weak prisoner's dilemma with $S=0$ and $k=0.1$. The behavior of WSLS is different from the imitation dynamics, especially above $T=1$, where here we have a non-zero cooperation level. Note also how the WSLS has a smooth drop in $T\simeq 0.6$.}
\label{fig_comp2D}
\end{figure}

We start by comparing the WSLS model to the usual imitation model for the weak prisoner's dilemma ($S=0$). Figure \ref{fig_comp2D} shows the fraction of cooperation in the equilibrium, $\rho$, as a function of the parameter T. In contrast to  the Imitation model, that exhibits a phase transition near $T_c=1.04$ where cooperators are extinct \cite{szabo_pre05,Szabo2007}, the WSLS model has a smooth drop in cooperation levels, but cooperation is not extinct, even for large $T$. This result agrees with the predictions of coexistence of cooperation found in our mean-field approximation.  Also notice that cooperation is smaller in the WSLS model for small values of $T$ compared to the imitation model. Recall that results for the imitation model are only for the sake of comparison in this work, since other papers study this model in depth (for a comprehensive review see \cite{Szabo2007}).

\begin{figure}
\centering{\includegraphics[width = 7.5cm]{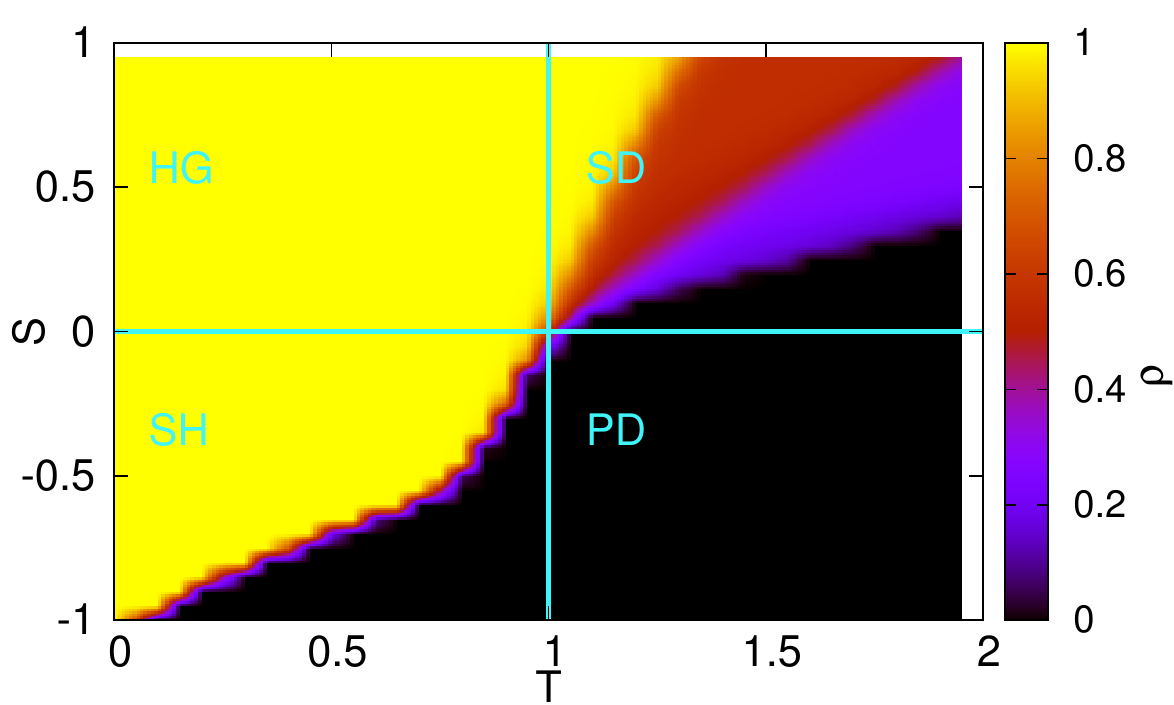}}

    \vspace{0.2cm}

\centering{\includegraphics[width = 7.5cm]{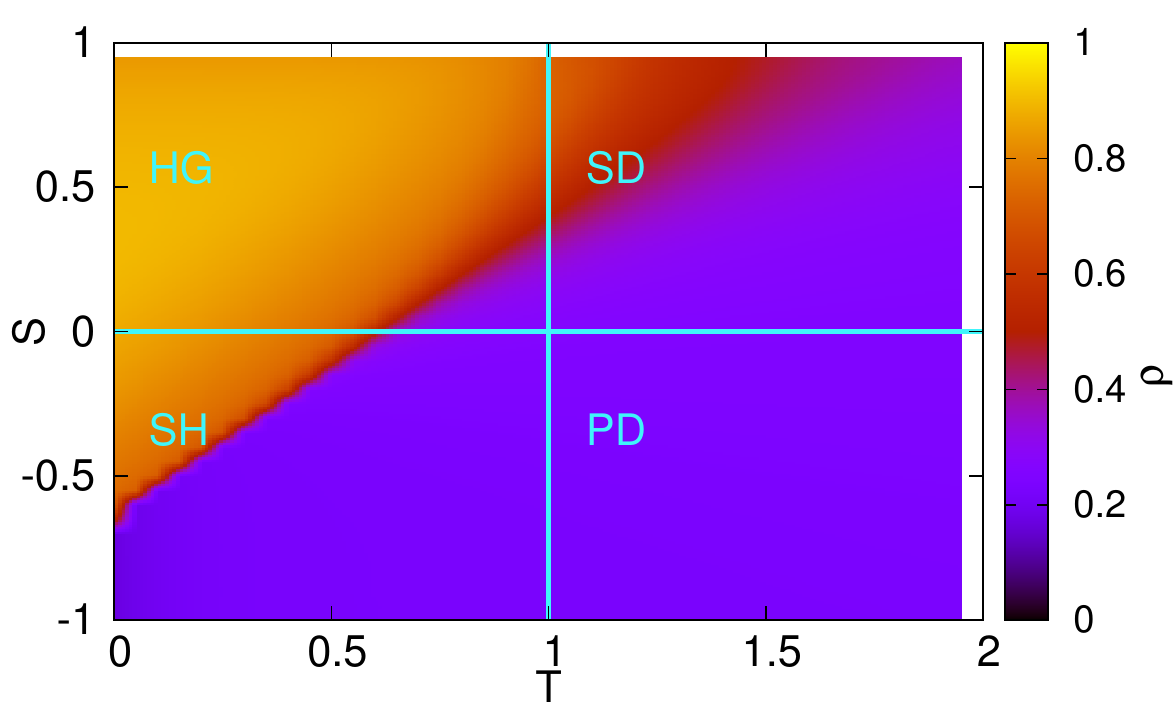}}
\caption{Phase diagram depicting the fraction of cooperators (color bar), depending on $S$ and $T$ for the two different models (WSLS and imitation).  In the imitation model (top), as expected, there is full cooperation in the harmony game quadrant, some coexistence in the SD quadrant, a sharp division in the SH quadrant, while most part of the PD quadrant is dominated by defection, with cooperators surviving only near the weak PD regime, with $T_c\approx1.04$. For the WSLS model (bottom) the cooperation is widespread everywhere, although lower than for imitation in the HG quadrant. We can see that the model is efficient in maintaining the co-existence of strategies even for high values of $T$.}
\label{fig_phase}
\end{figure}

Figure \ref{fig_phase} shows cooperation level in the entire  T-S plane for the imitation and the WSLS model, with cooperation represented by the color scale.

As expected, in the imitation model there is full cooperation in the HG quadrant, a mixture of full cooperation and coexistence in the SD quadrant, a sharp division of full cooperation or full defection for the SH quadrant and only defection for most of PD quadrant. The only coexistence in PD game is for small $S$ values (around $S=-0.01$) \cite{szabo_pre05}.

In the WSLS model we see a totally different behavior. Cooperation coexists with defectors in the entire phase diagram. More specifically,  cooperation is mostly enhanced in the HG quadrant; there is a sharp division in the SH quadrant; in the SD there is a smooth variation; and in the PD quadrant cooperation has the lowest values. The interesting result is that cooperation levels are non-zero for the whole phase diagram, the lowest value around $0.2$.

\begin{figure}
\centering{\includegraphics[width = 6cm]{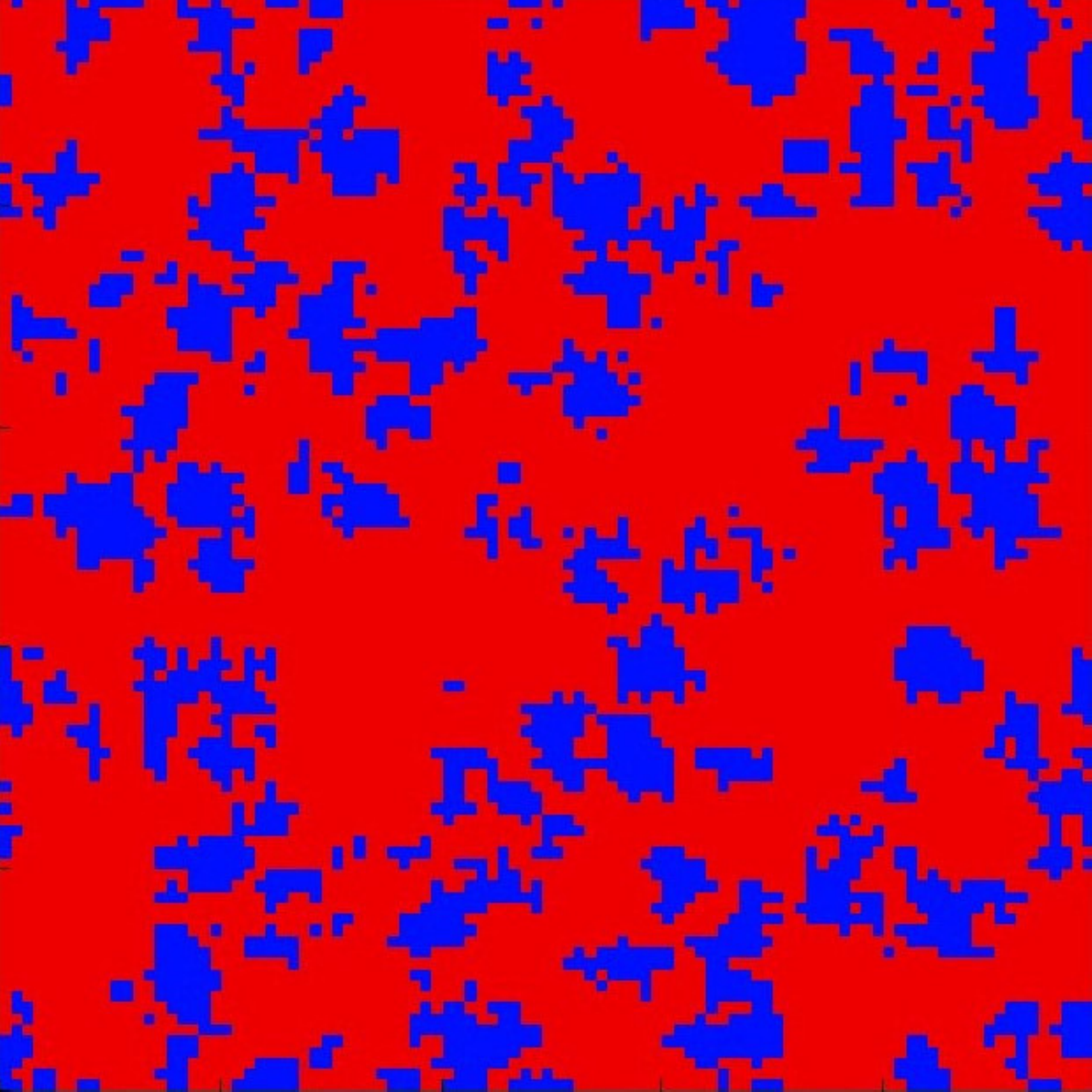}}

    \vspace{0.2cm}

\centering{\includegraphics[width = 6cm]{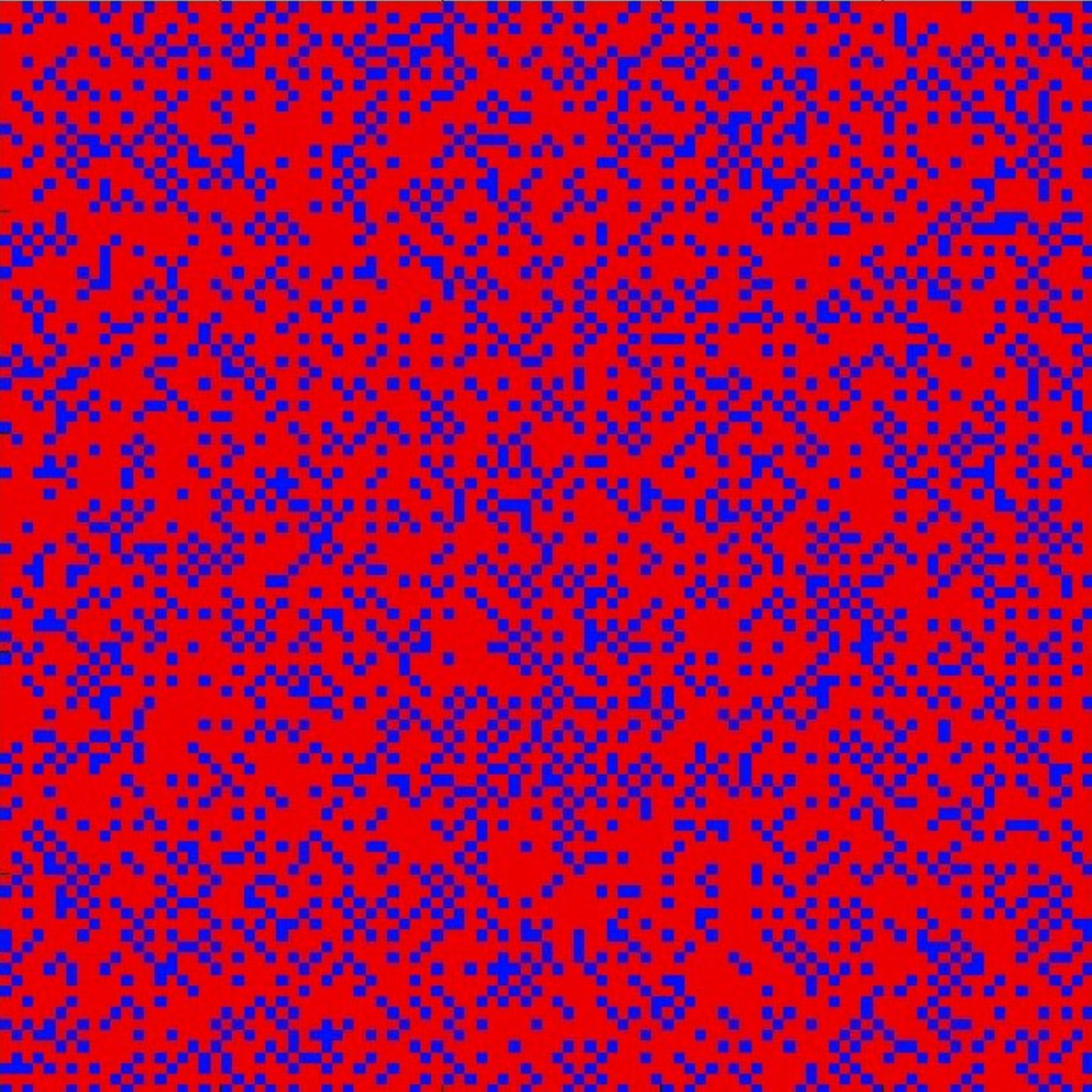}}
\caption{Typical snapshots from the Imitation (top) and WSLS (bottom) models, cooperators are dark blue and defector light red. We can see different spatial organization patterns that spontaneously emerge. We depict a typical PD game where $\rho \approx 0.25$ and $S=-0.01$ for both models. The temptation level is Imitation($T= 1.023$) and WSLS($T= 1.1$).}
\label{fig_pd}
\end{figure}

It is insightful to see one-frame snapshots of the square lattices after the system reaches a dynamical equilibrium. Figure \ref{fig_pd} shows  snapshots of the lattice for each model (both imitation and WSLS have the same fraction of cooperation and are playing the prisoner's dilemma). Note that the spatial organization of cooperators is totally different. While in the imitation model, cooperators form islands to survive (as expected) \cite{Szabo2007}, in the WSLS model cooperators and defectors are homogeneously distributed, forming a checkerboard-like pattern. Similar patterns for innovative dynamics were also found in \cite{Fort2005}. Remarkably, the weak correlation present in the WSLS is the cause of the success of our mean-field approximation.

To understand the microscopic mechanisms underlying the evolution of cooperation, we study the deterministic case, obtained in the limit $k \rightarrow \infty$ and, again, we set $S=0$ for simplicity. We stress that  simulations in this limit yield the similar results are for intermediate $k$ values.  The evolution of cooperation in the WSLS model, in contrast to the imitation model, must rely on a different microscopic mechanisms to promote cooperation, as indicated by the distinct spatial organizations of strategies. Since the statistical nature of the Fermi-Dirac distribution does not allow us to obtain a simple picture of the mechanisms, we will focus on the deterministic case obtained in the limit $k \rightarrow \infty$, where the site definitively changes its strategy if $u_i<\bar{u}$, or stays the same in the opposite situation. We set $S=0$, again, for simplicity.

\begin{figure}
\centering{\includegraphics[width = 8cm]{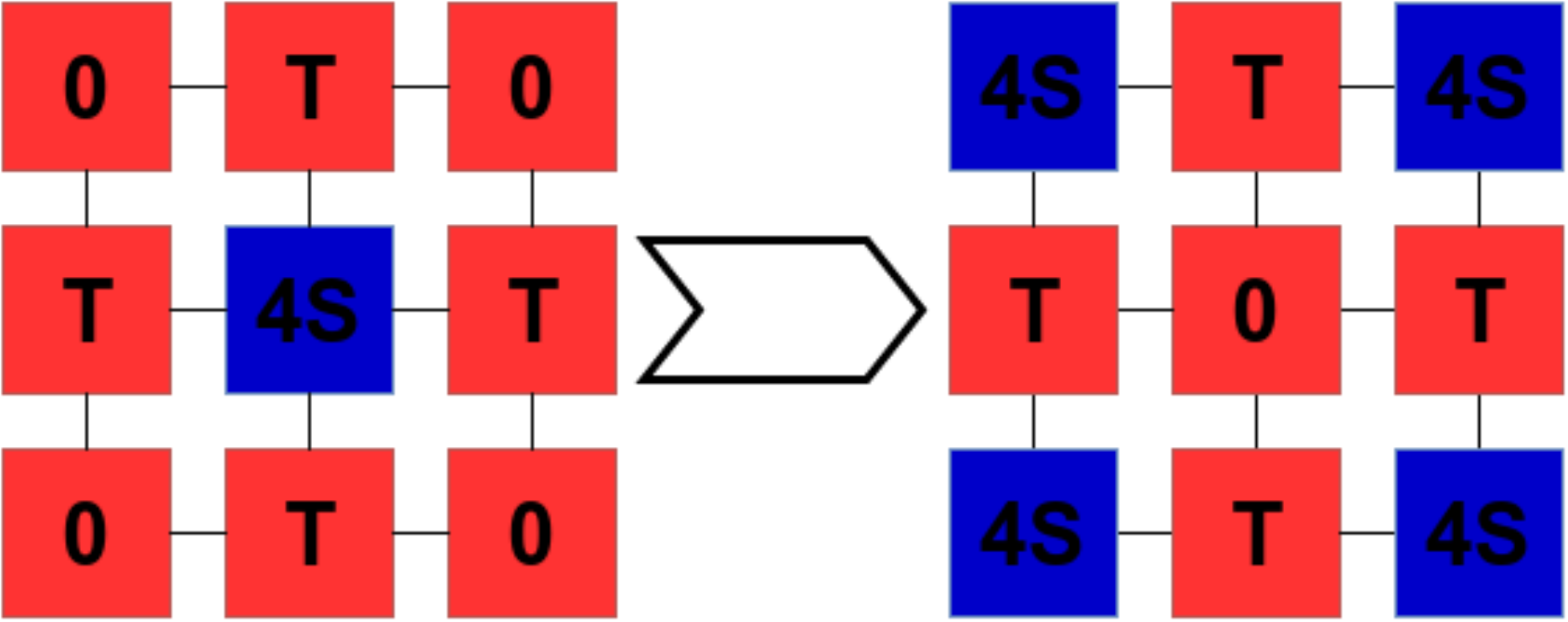}}
\caption{Cooperator (dark blue) surrounded by defectors (light red) in the deterministic version of the WSLS model. Although the focal site changes to defection, its second neighbors become cooperators due to the lower payoff when compared to the first neighbors.}
\label{fig_wslsblock}
\end{figure}

Analyzing a cooperator surrounded by defectors, we see that cooperation spreads to the second next neighbors, instead of to its first neighbors, as can be seen in figure \ref{fig_wslsblock}. Since the payoff of central site C is lower than the average payoff of its neighbors, the central site will change to a defector. But at the same time all second neighbor defectors have a payoff (zero) lower than the payoff ($T$)  of the first neighbors defectors, what causes  the second neighbors, to turn to cooperation. The basic mechanism is the greediness of defectors, surrounded by other defectors. This makes they constantly change strategy if there is at least one defector faring better.  In other worlds, the greediness of defectors, leads to their downfall. This micro-mechanisms also points to the curious phenomena that cooperators do not to stick together in this WSLS model.

Our results suggest that the effect of the WSLS mechanism on cooperation is not directly related to network reciprocity, where cooperators form clusters of cooperation that provide mutual help \cite{Nowak1992a, wardil_epl09, wardil_pre10, szolnoki_pre11c}. This can be tested by varying the network topology. We therefore investigate how the models behaves in scale-free networks \cite{Brown2004, Barabasi1999}, a well studied case of topology that enhances cooperation \cite{ohtsuki_jtb06, Santos2006, wardil_jpa11}. To have a robust result, we study both the absolute payoff (the payoff of a player is just the sum of payoffs obtained in each interaction) and the normalized payoff  (the absolute payoff divided by the number of neighbors) \cite{Masuda2007, Szolnoki2008a}. The  networks are generated with the Krapivsky-Redner algorithm \cite{krapivsky_pre01, Amaral2016}, a type of growing network with redirection (GNR) method. We used scale-free networks with $10^{4}$ nodes, irrationality $k=0.1$, weak PD ($S=0$) and average connectivity degree of $2.7$.

\begin{figure}
\centering{\includegraphics[width = 8cm]{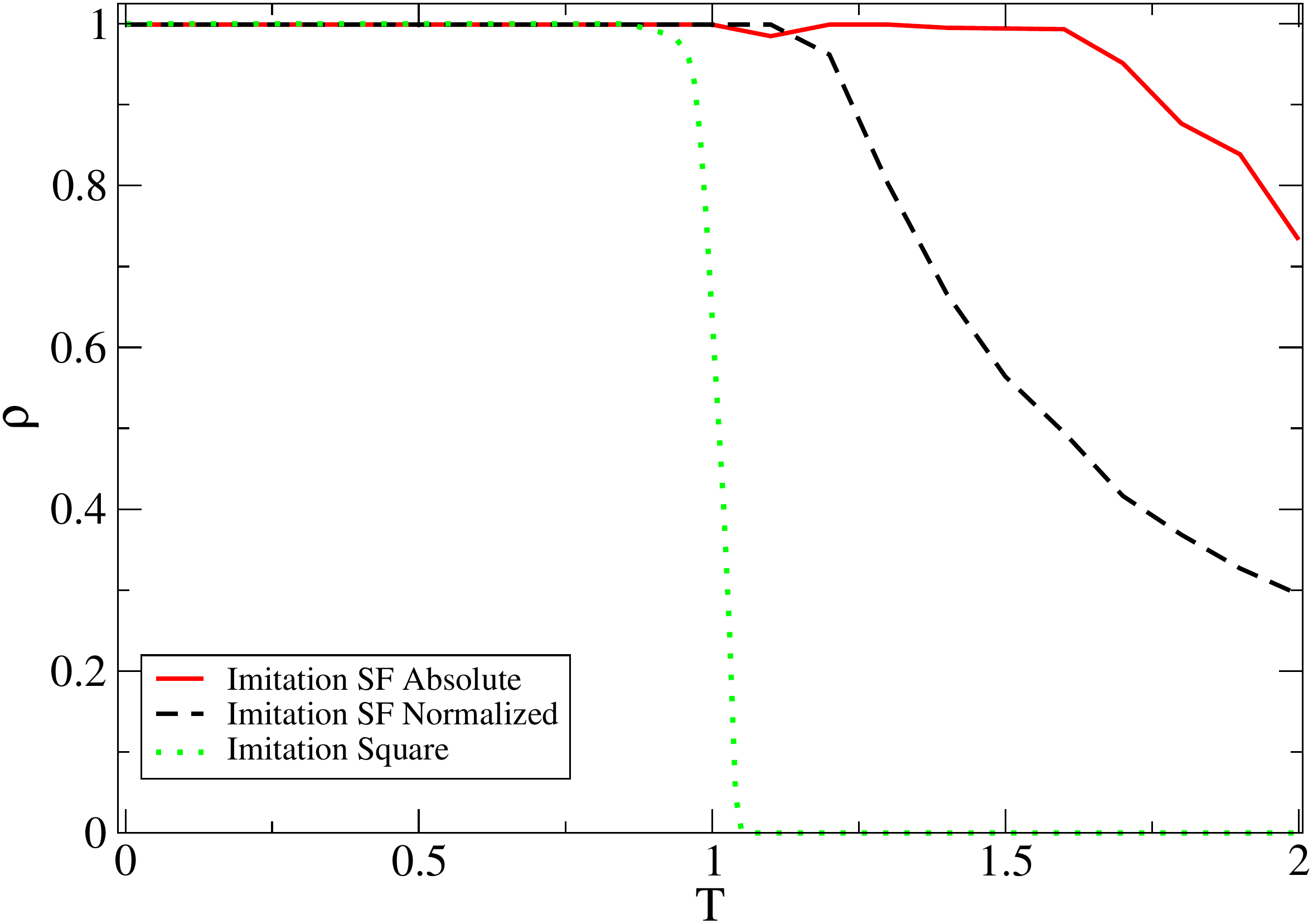}}

    \vspace{0.5cm}

\centering{\includegraphics[width = 8cm]{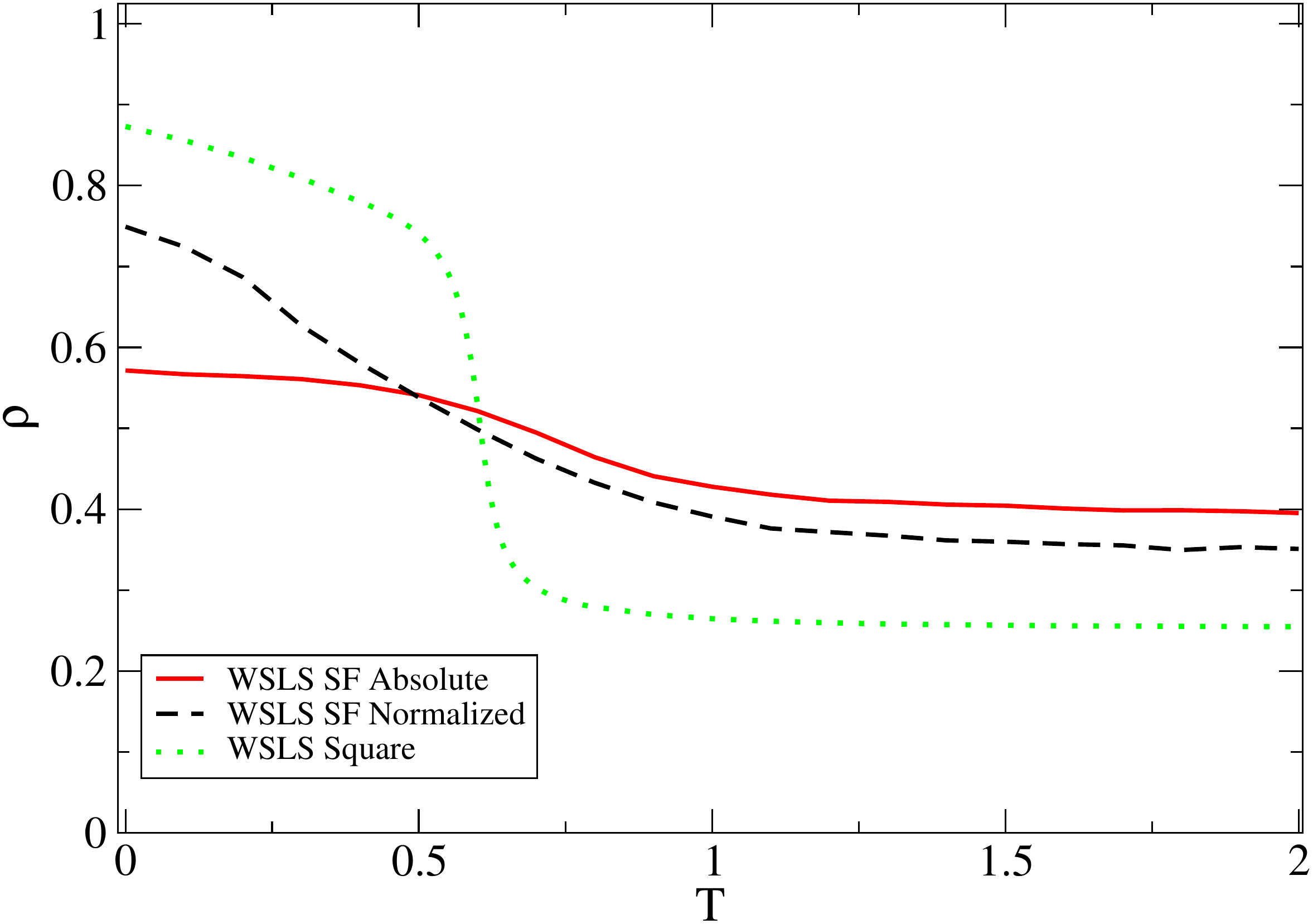}}
\caption{Fraction of cooperation as a function of $T$ in two different topologies: square lattice, and scale-free network with absolute and normalized payoff. We see for the imitation model (top) that the scale-free network with absolute payoff highly enhances cooperation, while the normalized payoff dampens this boost to a substantial degree. The lowest cooperation is obtained on the square lattice. For the WSLS (bottom) the different topologies have little effect on the evolution of cooperation.}
\label{fig_comp}
\end{figure}

In the imitation model, scale-free networks enhance cooperation when absolute payoffs are considered \cite{Masuda2007, Szolnoki2008a}. The enhancement is dampened if normalized payoff is used, but still scale-free topology favors cooperation more than  square lattices \cite{Szabo2007, santos_prsb06, nowak_aam90, Szolnoki2008a, Brown2004, Santos2006, santos_prl05, szabo_pre05, tang_epjb06, ohtsuki_jtb06, wardil_jpa11}. Figure \ref{fig_comp} shows the results for scale free network and square lattice, comparing both imitation and WSLS models.  The effect of scale-free topology in the promotion of cooperation is much weaker in the WSLS model than in the imitation model.

The study of the WSLS model in scale-free networks  indicates that  topology  has a small effect in the evolution of cooperation in the WSLS model, when compared to the effect it has in the imitation model. It is an interesting result, if we take into account that topology strongly affects imitative dynamics \cite{roca_pre09, santos_pnas06, Szabo2007, santos_prsb06, nowak_aam90, Szolnoki2008a, Brown2004, Santos2006, santos_prl05, szabo_pre05, tang_epjb06}. The topology independence strengths the point that the mechanisms promoting cooperation in the WSLS model relies on another source, other than spatial reciprocity. Recent works  found that the system dependency on topology can be irrelevant for some innovative update rules like best response, extortion, and myopic  \cite{roca_epjb09, szolnoki_pre14}.

Finally, our analysis reinforces the fact that  network reciprocity is dependent on the kind of strategy update rule that is used, and not on the mere presence of network structure. Indeed, the replicator equation is equivalent to the Monte Carlo dynamics only when individuals change strategy by copying each other \cite{Szabo2007, MaynardSmith1982a, roca_epjb09}, and this is not the case here. In the WSLS model, strategies are not replicated in the sense that they are transmitted from more successful individuals to less successful ones. Instead, the success of neighbors only influences ones decision on whether to keep or to change its own strategy.

\section{Conclusions}

In this work we studied the Win-Stay-Lose-Shift mechanism with local average aspiration in the evolutionary game framework using the master-equation and Monte Carlo analysis. The basic idea is that players  aspire to be at least as wealthy as the average of their neighbors, changing strategy otherwise. Cooperative behavior always face the challenge to survive in a population of self-interest individuals, since defecting against cooperators is more profitable.  However, we found that if the motivation of faring as good as the neighbors is the base of individual behavior, cooperation will emerge in coexistence with defection. This result was supported by computer simulations in the entire range of payoff parameters and was confirmed by mean-field approximations.

In imitation models, compact cooperative islands arise around seeds of successful cooperators. At the border between cooperators and defectors, the latter will do better and the islands will spread, sometimes shrink, and in general move across the network. In WSLS models, successful defectors will cause an erosion of compact defector patches, since  internal defectors will change their strategies due the high success of the defectors at the border. This drastically affects the whole population, causing cooperators to be homogeneously distributed in a checker-board like manner, instead of forming islands. This also results in cooperators lingering even for high values of temptation and, at the same time, defectors being always present in the population, even for strongly cooperative games like the harmony game.

The analytical predictions shows a minimum cooperation level above zero, even for high temptation. The stability was reached independent of initial conditions, and we prove that the ODE have a stable equilibrium point with $\rho^*>0$ for large $T$. We tested the model using asynchronous Monte Carlo dynamics in square and scale-free lattices. Using numerical simulations for Monte Carlo we still found the basal cooperation level and independence with the initial state. Even more, cooperation is widespread through the entire $T-S$ diagram, differently from the classical non-innovative dynamics where cooperation does not linger on the prisoners dilemma for most values of $T$ and $S$. We deeply analyzed the microscopical mechanism that leads to the support of cooperation using deterministic dynamics. We found out that in this innovative process, cooperation is transferred to the second neighbors, instead of the first ones as in copy mechanisms. This drastically affects the whole population, causing cooperators to be homogeneously distributed, instead of forming islands. This also results in cooperation lingering for high values of temptation. At the same time defectors are always present, even in strongly cooperative games like the harmony game. We studied the model on scale-free networks and found that the classical result of cooperation enhancement due to network reciprocity remained absent, further supporting the claim that innovative dynamics does not rely on such reciprocity to maintain cooperation. This is interesting, also in the light of recent research on the importance of the integration of cognitive abilities in game theoretical models \cite{bear_pnas16}, and the fact that human cooperation is likely more related to cognitive strategies than to effects stemming from replicator dynamics.

Lastly, our work  highlights the relevance of the proper choice of the updating rule when modeling human behavior. While the evolution of strategies in simpler animals over long time scales can be described by the replicator dynamics, it is not always the case when individuals have higher cognitive capacity and can make choices very fast, in time scales that are much shorter than the typical time to induce an evolutionary transition. We note that our results support preceding research on innovative dynamics, fast decision making, and intuitive cooperation \cite{rand_nc14, capraro2014heuristics, capraro2015social, capraro2016heuristics}, highlighting also the importance of the different updating rules. In this sense it becomes clear that one should be very careful when choosing a model to describe a real-life situation. We hope that this paper will motivate further research along this line in the near future.

\begin{acknowledgments}
This research was supported by the Brazilian Research Agencies CAPES- PDSE (Proc. BEX 7304/15-3), CNPq and FAPEMIG, and by the Slovenian Research Agency (Grants J1-7009 and P5-0027).
\end{acknowledgments}

\end{document}